\documentclass[b5paper,10pt,openright]{book}  

\usepackage{times,amsfonts}
\usepackage{psfig}
\usepackage{epsf}
\usepackage{rotating}
\usepackage{wrapfig}
\usepackage{graphicx,times}
\usepackage{graphics}
\usepackage{setspace,dropping,section,relsize}
\usepackage{caption2,topcapt}
\usepackage[Lenny]{fncychap}
\usepackage{amsmath}
\usepackage[dutch,english]{babel}
\usepackage{amssymb,latexsym,stmaryrd}
\usepackage[innercaption]{sidecap}
\usepackage{natbib,natbibspacing}
\usepackage{multirow}
\bibpunct{(}{)}{;}{a}{}{,}
\usepackage{fancyheadings}
\renewcommand{\chaptermark}[1]%
   {\markboth{\MakeUppercase{\chaptername}\thechapter.\ #1}{}}

\renewcommand{\captionlabeldelim}{--.}
\renewcommand{\captionfont}{\small}
\renewcommand{\captionlabelfont}{\normalsize \bf}

\newcounter{savesection}
  {\setcounter{savesection}{\value{section}}%
   \setcounter{section}{0}%
   \renewcommand\thesection{\thechapter.\Alph{section}}}%
  {\setcounter{section}{\value{savesection}}}
\def\etal{{\it et al.}}
\def\lae{\mathrel{<\kern-1.0em\lower0.9ex\hbox{$\sim$}}}
\def\gae{\mathrel{>\kern-1.0em\lower0.9ex\hbox{$\sim$}}}

\def\kms{km~s$^{-1}$}

\def\rO3HB{$[$OIII$]$5007\slash H$\beta$~}

\def\rN2HA{$[$NII$]$6583\slash H$\alpha$~}
\def\rS2HA{$[$SII$]$\slash H$\alpha$~}

\newcommand{\hi}{\ion{H}{i}}

\newcommand{\oiii}{[\ion{O}{iii}]}

\DeclareMathAlphabet{\mathsc}{OT1}{cmr}{m}{sc}
\def\testbx{bx}%
\DeclareRobustCommand{\ion}[2]{%
\relax\ifmmode
\ifx\testbx\f@series
{\mathbf{#1\,\mathsc{#2}}}\else
{\mathrm{#1\,\mathsc{#2}}}\fi
\else\textup{#1\,{\mdseries\textsc{#2}}}%
\fi}

\usepackage{natbib}

\begin{document}

\newsavebox{\rotbox} 

\pagestyle{empty}

\pagestyle{empty}

\begin{figure}[h]
\begin{center}
\epsfxsize=1.1cm
\epsfbox{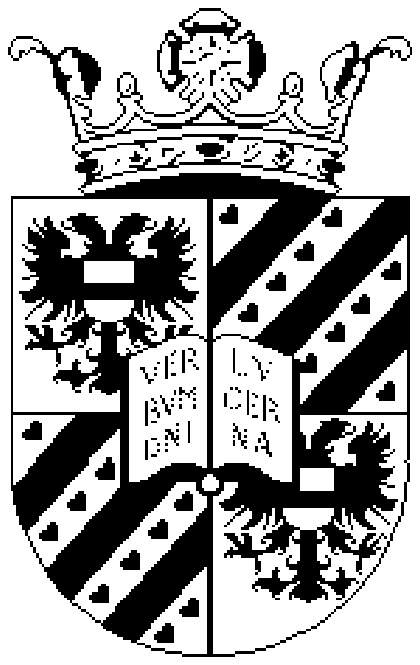}
\end{center}
\end{figure}
\begin{center}
\large
{\sc \Large Rijksuniversiteit Groningen}\\
\vspace{2.0cm}
{\hspace{-0.0cm} \bf \LARGE
Host galaxies and environments of\strut\\
compact extragalactic radio sources\strut}\\
\vspace{1.5cm}
{\Large \sc Proefschrift\strut}\\
\smallskip
ter verkrijging van het doctoraat in de\strut\\
Wiskunde en Natuurwetenschappen\strut\\
aan de Rijksuniversiteit Groningen\strut\\
op gezag van de\strut\\
Rector Magnificus, dr.\ F.\ Zwarts,\strut\\
in het openbaar te verdedigen op\strut\\
vrijdag 24 februari 2006\strut\\
om 10:30 uur\\
\vspace{3.0cm}
door\\
\bigskip
{\bf \'Alvaro Labiano Ortega}\\
\bigskip
geboren op 14 januari 1977\\
\smallskip
te Madrid, Spanje\\
\end{center}

\vfill
\newpage

{\large
\noindent
\begin{tabular}{ll}
  Promotor:              & Prof. Dr. P. D. Barthel\strut\\
  Co-promotores:           & Dr. C. P. O'Dea\\
                   & Dr. R. C. Vermeulen\\
  	& \\
  Beoordelingscommissie: & Prof. Dr. A. G. de Bruyn\\
                         & Prof. Dr. J. M. van der Hulst\\
                         & Prof. Dr. R. F. Peletier\\
\end{tabular}
}

\vspace{15cm}
\noindent
{\large \bf ISBN-nummer: 90-9020416-4}

\pagestyle{empty}

\null

\vspace{18cm}

\begin{flushright}
{\it \Large

To my parents
}
\end{flushright}
\pagestyle{fancyplain}  
\renewcommand{\thepage}{\roman{page}}

\selectlanguage{english}

\chapter{Thesis Outline}
\label{chapter_intro}
\markboth{\textsc{Chapter 1:} Thesis Outline}{}
\thispagestyle{empty}
\sloppy 

\renewcommand{\thepage}{\arabic{page}}
\setcounter{page}{1}


\vspace*{-7.5cm}
\begin{quote}
 \hspace{8.0cm}
 {\it \small \raggedright Though this be madness,\\
   \hspace{8.0cm}  yet there is method in 't.\\ 
   \vspace{0.5cm}
   \hspace{8.7cm}  {\scriptsize \sf William Shakespeare}\\
   \hspace{9.7cm}  {\scriptsize \sf Hamlet, II, 2.}}\\
\end{quote}

\vspace{5.5cm}


\section{Thesis project}

By the beginning of this thesis, GPS and CSS sources stood as different classes of objects by themselves, and a lot of research had just been carried out. However, there were still questions to be answered. In my opinion, the two main questions concerning GPS and CSS sources were:

\begin{enumerate}
\item[1)]{Where do they stand in relation with the other radio sources and AGN?}

\item[2)]{How do GPS/CSS interrelate with their host?} 

\end{enumerate}

Both questions had been around for some time, however, the first one was sooner to be answered. There were two important workshops held on 1999 ({\it Life cycles of radio galaxies}\footnote{Proceedings published in New Astronomy Reviews, editors: J.A. Biretta, A.M. Koekemoer, E.S. Perlman, C.P. O'Dea. May 2002, Vol. 46, Nos. 2--7}) and 2002 ({\it The Third Workshop on CSS and GPS  radio sources}\footnote{Proceedings published by PASA, 2003, Vol. 20, editors: T. Tzioumis, W. de Vries, A.M. Koekemoer, I. Snellen}). In the first one, participants showed growing evidence supporting the young scenario. By the end of the 2002 workshop, most participants were convinced that GPS and CSS sources are (or seem to be) young.\\

Research had been carried out focussing on the interrelation between the radio source and the host. However it was a by far less active field. The main developments on this subject were done on alignment effect of the radio source and the line emission gas. This thesis takes it from there and advances in our understanding of the interrelation between the radio source and the host and how it is affecting both. \\  

Other issues addressed in the thesis:

\begin{enumerate}

\item[3)]{Mapping of compact radio sources.} 

\item[4)]{Enlargement and improvement of GPS and CSS samples.} 

\item[5)]{Properties of the gas and stellar population of the host.}  

\item[6)]{Ionization mechanisms of the gas in the host.} \\  

\item[7)]{Gas content and distribution in the host.}  

\item[8)]{First near-UV study of GPS and CSS sources.}  

\item[9)]{AGN--starburst connection.}  

\end{enumerate}

\section{Thesis outline}

Chapter 2 presents radio maps of compact sources in the Southern hemisphere. The data are being compared with other measurements to classify and look for unidentified GPS and CSS sources in the sample. Number statistics of GPS and CSS quasars and galaxies are presented and compared with the Northern hemisphere. \\

Chapter 3 presents optical observations of hosts of candidate GPS radio sources. Identification and spectroscopy of the hosts are carried out, aiming to complete the GPS identification and redshift entries of the O'Dea91 \etal GPS master list. New identifications, magnitudes, redshifts and emission line gas properties are reported. Models of stellar populations are compared with the observations. Literature data are collected to study previous GPS identifications.\\

Chapter 4 presents Hubble Space Telescope long slit spectroscopy of CSS sources, studying the ionization mechanisms of the emission line gas, as well as its properties. All sources show a combination of photoionization and --fast-- shock ionization. The expansion of the radio source is clearly affecting the optical emission properties of the host. \\ 

Chapters 5 and 6 present high resolution radio imaging and spectroscopy of two CSS sources and a GPS source, and study the contents, distribution and properties of cold gas in the host, revealing that it seems to be associated with the emission line gas and presenting a completely  new interpretation of an old known GPS source.\\

Chapter 7 presents the first study of the near-UV emission in hosts of GPS and CSS sources. Alignment between the radio source and UV emission of the host is found, as well as small UV emitting regions, consistent with recent star formation. A comparative study with large radio sources is presented. The connection between bursts of star formation and the presence of a powerful radio source is explored.\\

Chapter 8 summarizes the main results of this thesis and discusses its implications. It also describes the work carried out by other groups concerning GPS and CSS sources and attempts to make a coherent picture. The last section of the chapter suggests possible lines of future research.

 \markboth{}{} 				

\chapter{Compact and extended radio sources in the southern sky}
\label{chapter_2}
\markboth{\textsc{Chapter 2:} Compact and extended radio sources in the southern sky}{}
\thispagestyle{empty}
\sloppy 

\begin{minipage}[c]{130mm}
\begin{center}
\normalsize
{\it Preliminary version of a paper to be co-authored by:} \\ 
A. Labiano, P.D. Barthel, R.W. Hunstead, R.T. Schilizzi, J. Bland-Hawthorn

\end{center}
\end{minipage}

\vspace{1cm}


\begin{center}
\begin{minipage}[c]{130mm}
  
\dropping[0pt]{2}{W}\textsc{e} present Very Large Array (VLA) observations of a subset of southern extragalactic sources selected from the Molonglo Southern 4 Jy (MS4) Sample, demonstrating that MS4 is comparable to the northern Third Cambridge Revised catalogue. Several interesting individual objects are discussed, including new galactic sized CSS objects. New CSS sources are found: B0614--349, B0615--365, B0646--398, B0707--359, B1015--314, B2259--375 and B2339--353 as well as a CSS candidate in the core of B0618--371. \\ 

\end{minipage}
\end{center}

 \markboth{}{}		

\chapter{Optical imaging and spectroscopy of candidate GPS radio sources}
\label{chapter_3}
\markboth{\textsc{Chapter 3:} Optical imaging and spectroscopy of candidate GPS radio sources}{}
\thispagestyle{empty}
\sloppy 

\begin{minipage}[c]{130mm}
\begin{center}
\normalsize
A. Labiano, P.D. Barthel, C.P. O'Dea, W.H. de Vries, I. P\'erez, S.A. Baum

\end{center}
\end{minipage}

\vspace{1cm}


\begin{center}
\begin{minipage}[c]{130mm}

\dropping[0pt]{2}{D}\textsc{eep} optical imaging and spectroscopy, obtained with the Very Large Telescope (VLT), is presented, targeting the host galaxies of candidate GPS radio sources from the master list of \cite{O'Dea91}. Our goal is to measure new redshifts, identify optical counterparts and address uncertain identifications. We measure redshifts for B0316+161, B0407--658, B0904+039, B1433--040, and identify the optical counterparts of B0008--421, B03161+161, B0407--658, B0554--026, B0742+103, B0904+039. We find that the previous identification for B0914+114 is incorrect. Using literature data we furthermore show that the following sources are not GPS: B0407--658, B0437--454, B1433--040, B1648+015.\\

\end{minipage}
\end{center}

 \markboth{}{}		

\chapter{HST/STIS low dispersion spectroscopy of three CSS sources: Evidence for jet-cloud interaction}
\label{chapter_4}
\markboth{\textsc{Chapter 4:} HST/STIS low dispersion spectroscopy of three CSS sources: \\ 
Evidence for jet-cloud interaction}{}
\thispagestyle{empty}
\sloppy 

\begin{minipage}[c]{130mm}
\begin{center}
\normalsize
A. Labiano, C.P. O'Dea, R. Gelderman, W.H. de Vries, D.J. Axon, P.D. Barthel, S.A. Baum, A. Capetti, R. Fanti, A.M. Koekemoer, R. Morganti, C.N. Tadhunter.

\textsc{Astronomy \&\ Astrophysics, 436, 493 (2005)}
\end{center}
\end{minipage}

\vspace{1cm}


\begin{center}
\begin{minipage}[c]{130mm}
  
\dropping[0pt]{2}{W}\textsc{e} present Hubble Space Telescope Imaging Spectrograph long-slit spectroscopy of the emission line nebulae in the compact steep spectrum radio sources 3C~67, 3C~277.1, and 3C~303.1.  We derive BPT \citep[Baldwin- Philips-Terlevich;][]{Baldwin81} diagnostic emission line ratios for the nebulae which are consistent with a mix of shock excitation and photo-ionization in the extended gas. In addition, line ratios indicative of lower ionization gas are found to be associated with higher gas velocities.  The results are consistent with a picture in which these galaxy scale radio sources interact with dense clouds in the interstellar medium of the host galaxies, shocking the clouds thereby ionizing and accelerating them.

\end{minipage}
\end{center}

 \markboth{}{} 		

\chapter{\ion{H}{i} absorption in 3C~49 and 3C~268.3}
\label{chapter_5}
\markboth{\textsc{Chapter 5:} \ion{H}{i} absorption in 3C~49 and 3C~268.3}{}
\thispagestyle{empty}
\sloppy 

\begin{minipage}[c]{130mm}
\begin{center}
\normalsize
{\it Accepted for publication in A\&A. In press} \\ 
A. Labiano, R.C. Vermeulen, P.D. Barthel, C.P. O'Dea, J.F. Gallimore, S.A. Baum, W.H. de Vries

\end{center}
\end{minipage}

\vspace{1cm}


\begin{center}
\begin{minipage}[c]{130mm}
  
\dropping[0pt]{2}{W}\textsc{e} present and discuss European VLBI Network UHF band spectral line observations, made to localise the redshifted 21cm \ion{H}{i} absorption known to occur in the subgalactic sized compact steep spectrum galaxies 3C~49 and 3C~268.3. We have detected \ion{H}{i} absorption towards the western radio lobe of 3C~49 and the northern lobe of 3C~268.3. However, we cannot rule out the presence of similar amounts of \ion{H}{i} towards the opposite and much fainter lobes. The radio lobes with detected \ion{H}{i} absorption (1) are brighter and closer to the core than the opposite lobes; (2) are depolarized; and (3) are associated with optical emission line gas. The association between the \ion{H}{i} absorption and the emission line gas, supports the hypothesis that the \ion{H}{i} absorption is produced in the atomic cores of the emission line clouds. Our results are consistent with a picture in which compact steep spectrum sources interact with clouds of dense gas as they propagate through their host galaxy. We suggest that the asymmetries in the radio and optical emission can be due to interaction of a two sided radio source with an asymmetric distribution of dense clouds in their environment
\end{minipage}
\end{center}

 \markboth{}{} 		

\chapter{\ion{H}{i} in the one-sided ``compact double'' radio~galaxy B2050+364}
\label{chapter_6}
\markboth{\textsc{Chapter 6:} \ion{H}{i} in the one-sided ``compact double'' radio~galaxy B2050+364}{}
\thispagestyle{empty}
\sloppy 

\begin{minipage}[c]{130mm}
\begin{center}
\normalsize
{\it Accepted for publication in A\&A. In press} \\ R.C. Vermeulen, A. Labiano, P.D. Barthel, C.P. O'Dea, J.F. Gallimore,S.A. Baum, W.H. de Vries

\end{center}
\end{minipage}

\vspace{1cm}


\begin{center}
\begin{minipage}[c]{130mm}
  
\dropping[0pt]{2}{E}\textsc{uropean} VLBI Network spectral imaging of the ``compact
double'' radio source B2050+364 in the UHF band at 1049~MHz
has resolved the \hi\ absorbing region, and has shown a faint
continuum component to the North (N), in addition to the well-known
East-West double (E, W).\\

Re-examination of VLBI continuum images at multiple frequencies
suggests that B2050+364 may well be a one-sided core-jet
source, which appears as a double over a limited frequency
range. One of the dominant features, W, would then be the innermost
visible portion of the jet, and could be at or adjacent to the
canonical radio core. The other, E, is probably related to shocks at a
sudden bend of the jet, towards the extended steep-spectrum region N.\\

A remarkably deep and narrow \hi\ absorption line component extends
over the entire projected extent of B2050+364.  It
coincides in velocity with the \oiii\ optical doublet lines to within
10~\kms.  This \hi\ absorption could arise in the atomic cores of NLR
clouds, and the motion in the NLR is then remarkably coherent both
along the line-of-sight and across a projected distance of $>300$~pc on
the plane of the sky.\\

Broader, shallower \hi\ absorption at lower velocities covers only the
plausible core area W\null. This absorption could be due to gas
which is either being entrained by the inner jet or is flowing out from
the accretion region; it could be related to the BLR.\\

\end{minipage}
\end{center}

 \markboth{}{} 		

\chapter{Star formation in hosts of compact radio galaxies}
\label{chapter_7}
\markboth{\textsc{Chapter 7:} Star formation in hosts of compact radio galaxies}{}
\thispagestyle{empty}
\sloppy 

\begin{minipage}[c]{130mm}
\begin{center}
\normalsize
A. Labiano, C.P. O'Dea, P.D. Barthel, W.H. de Vries, S. A. Baum

\end{center}
\end{minipage}

\vspace{1cm}


\begin{center}
\begin{minipage}[c]{130mm}

\dropping[0pt]{2}{W}\textsc{e} present near ultraviolet imaging with the Hubble Space Telescope Advanced Camera for Surveys, targeting young radio galaxies (Gigahertz Peaked Spectrum and Compact Steep Spectrum sources), in search of star formation regions in their hosts. We find near UV light which could be the product of recent star formation in eight of the nine observed sources. Stellar synthesis models are consistent with a burst of star formation before the formation of the radio source. However, observations at other wavelengths and colors are needed to definitively establish the nature of the observed UV light. In the CSS sources B1443+77 and B1814--637 the near UV light is aligned with and is co-spatial with the radio source. We suggest that in these sources the UV light is produced by star formation triggered and/or enhanced by the radio source. 

\end{minipage}
\end{center}

 \markboth{}{}		

\chapter{Concluding remarks}
\label{chapter_implicat}
\markboth{\textsc{Chapter 8:} Concluding remarks }{}
\thispagestyle{empty}
\sloppy 



\vspace*{-7.5cm}
\begin{quote}
 \hspace{6.0cm}
 {\it \small \raggedright With me poetry has not been a purpose, \\
   \hspace{9.3cm}  but a passion.\\ 
   \vspace{0.5cm}
   \hspace{9.2cm}  {\scriptsize \sf Edgar Allan Poe}\\
   \hspace{8.7cm}  {\scriptsize \sf }}\\
\end{quote}

\vspace{5.5cm}


\dropping[0pt]{2}{I}\textsc{n} this final chapter of the thesis, I summarize what we have learned about GPS and CSS sources in the last years. I list a few of the most (in my opinion) important questions in the field and highlight where this thesis has contributed (Section \ref{developments}). I summarize the most important results of each chapter and what I think is the main contribution of the thesis to the field (Sections \ref{chapters} and \ref{bottomline}). I also describe some of the ongoing work in our group and suggest possible lines of future research (Section \ref{future}).\\

\section{Developments in the field} 
\label{developments}

In the introduction of the thesis (Section 1.7), I listed some of the questions and issues concerning compact extragalactic radio sources that had yet to be answered by the time I embarked on the project. In this section I describe the latest developments in the field and highlight the contributions from this thesis. 

\begin{itemize} \item 
{\it Are GPS and CSS sources young?} 

The nature of GPS and CSS sources and their relation with larger radio sources has been an active issue for over thirty years \citep[e.g.,][]{Blake70, Breugel88}. In the past years \citep[e.g.,][]{Fanti95,Readhead96a}, growing evidence supporting the young scenario has been found. Spectral aging, energy supply arguments and, especially, measurements of expansion velocities support the young scenario \citep[e.g.,][]{Conway02, Polatidis03, Murgia03, Siemiginowska05}. 

However, not all GPS and CSS sources must be necessarily young. \cite{Stanghellini05} find large scale diffuse radio structures in GPS sources, which could be the relic of a past nuclear activity phase, suggesting a recurrent source. \cite{Marecki03} and \cite{Marecki05} find GPS and CSS sources where the activity has recently stopped, so even though they are young, they are --as they refer to them-- {\it  dying} sources. \cite{Siemiginowska03} observed a sample of GPS sources in X-rays and found that some\footnote{Some X-ray observations find that the hot gas cannot confine the GPS source \citep[e.g.,][]{O'Dea96b, O'Dea00}.} of their hosts may have enough gas to confine them. \\


\item
{\it How do GPS and CSS sources evolve?}   

According to the {\it young scenario}, GPS evolve into CSS and these into FR sources. However, it is not completely clear how this evolution takes place or if they evolve into FR I or FR II sources. Based on morphological similarities, it has been traditionally thought that they evolve into FR II. However there is no strong evidence supporting this (nor denying it) and there are models where GPS and CSS could evolve into FR I, both directly \citep[especially if the CSS is weak, e.g.,][]{Marecki05, Snellen00} or through a transient FR II phase \citep[e.g.,][]{Ghisellini01}. 

According to the general {\it classic} model, GPS/CSS grow into FR sources in a self-similar way, at roughly a constant velocity through an ambient medium which declines in density  as $\rho(R) \propto R^{-2}$ while the sources decline in radio luminosity as $L_{rad} \propto R^{-0.5}$ \citep[e.g.,][]{Fanti95,Begelman96,Readhead96b}. Most of the models developed later have been based on this one, including free-free or synchrotron self-absorption and varying the jet or the host physical parameters \citep[e.g.,][]{Kaiser97b, Bicknell03}. 

Modeling how the expansion takes place seems to be a matter far from solved. A few examples: \cite{Snellen00} introduced a variation to the {\it classic} model where GPS would increase in luminosity as they grow into CSS, which would then dim as they expand. With this variation they could explain the observed redshift distribution of GPS/CSS. A few years later, \cite{Tinti05} describe a model that predicts that same distribution with redshift but without the increase in luminosity in the GPS to CSS phase. Another matter of discussion is the self-similar expansion. It seems a little {\it ad hoc} to require the jet to expand through and interact with the host while maintaining geometric relationships. Recent models \citep[e.g.,][]{Jeyakumar05, Saxton05} include these interactions, and some of them obtain better predictions without including self-similar expansion \citep[e.g.,][]{Carvalho03}. However, \cite{Tschager03} present a new sample of weak CSS sources and argue that self-similarity is an essential, intrinsic characteristic of the expansion. 

The interest in modeling the expansion the radio source keeps growing and it is becoming an extremely active field. On the other hand, considerable amounts of work are being carried out with the purpose of increasing, improving and homogenizing the GPS and CSS samples, which will help addressing the problem and improving the models.\\

\item 
{\it Free-free of synchrotron self-absorption?}

This has been a less active field but it is still surrounded by controversy. The main trend is that the peak in the radio spectra of GPS and CSS sources is due to synchrotron self-absorption \citep[SSA, e.g.,][]{Snellen00}. However, even in sources where SSA seems to cause the peak, free-free absorption cannot completely be ruled out \cite[e.g.,][]{Tingay03}. Furthermore, there are sources where the turnover seems to be produced by free-free absorption \cite[FFA, e.g.,][]{Marr01, Shen05} and some models are sill developed based on FFA \citep[e.g.,][]{Kameno03}.\\

\item
{\it GPS quasars}

The problem with the nature of GPS quasars is still open. The majority of the sources that are currently classified as GPS quasars could be a different type of object, unrelated to compact, {\it real}, GPS/CSS sources \citep[e.g.,][]{Snellen99, Stanghellini03, Bai05}. A detailed description of the problem is given is Section 1.6 of the Introduction of this thesis. See also \cite{Snellen97}. Most of the work carried out trying to solve this problem has focused on observing known old samples of GPS sources with better telescopes and obtaining new samples. So far, the most useful techniques for improving the samples seem to be high resolution imaging and, particularly, searches for radio spectral variability \citep[e.g.,][]{Dallacasa00, Jauncey03, Tinti05b}. \\

\item
{\it Samples of GPS and CSS sources}

Addressing most, if not all, the above issues requires good samples. During the first years of GPS/CSS research, these sources were drawn from surveys and samples targeting general radio sources (3CR for example), with selection criteria that would not necessarily apply to GPS/CSS sources. However, this has changed in the last few years and there have been projects aimed at improving old samples \citep[e.g.,][; {\bf Chapters 2 and 3} of this thesis]{Vries95, Stanghellini98, Snellen98, Vries00b, Xiang02, Rossetti05} and obtaining new ones, targeting only (or mostly) GPS and CSS sources  \cite[e.g.,][]{Dallacasa00, Fanti01,Taylor03, Edwards04}. As samples get improved, enlarged and homogenized, progress is made in almost every field of research concerning GPS and CSS sources \citep[e.g.,][]{Tschager03, Marecki05, Stanghellini05, Tinti05}.\\

\item
{\it Cold gas in GPS and CSS sources}

Many GPS, CSS and FR sources are in interacting or merging systems \citep[e.g.,][]{Vries00, Johnston05}, therefore we expect their hosts to have dense nuclear environments. Furthermore, most of the radio loud AGN seem to live in elliptical galaxies \citep[e.g.,][]{Martel99, Dunlop03}. \cite{Walsh89} found that the presence of gas (and dust) in early type galaxies is correlated with the occurrence and strength of a central radio source. This gas is found both in molecular and atomic form \citep[e.g.,][]{Knapp96, Oosterloo99}. Observations at different wavelengths find central structures of gas in AGN hosts \citep[e.g.,][]{Gorkom89, Verdoes99, Evans99, Langevelde00, Morganti01}. If the gas is centrally concentrated and GPS/CSS sources are precursors (smaller versions) of FR sources, we expect GPS/CSS to show higher densities in the gas surrounding their radio lobes\footnote{However, extremely high densities may confine the sources.}. In 2003, using the recently improved Westerbork Synthesis Radio Telescopes, \cite{Vermeulen03} and \cite{Pihlstrom03} presented a study of the occurrence and properties of atomic gas associated with compact radio sources. They found that $\sim50\%$ of GPS and CSS sources showed \ion{H}{i} 21 cm absorption, in contrast to normal elliptical galaxies \citep[$\lesssim 10\%$, ][]{Gorkom89}. Furthermore, they found that GPS sources tend to have higher \ion{H}{i} column densities than CSS sources and that these densities were consistent with the young scenario. However, their data lack spatial resolution to accurately locate the \ion{H}{i} absorption. Just a few high resolution \ion{H}{i} observations of GPS/CSS sources were available at that time \citep[e.g.,][]{Conway96, Peck98, Peck99, Peck02}. We are involved in a project aimed at obtain high resolution \ion{H}{i} observations with the European VLBI Network. {\bf Chapters 5 and 6} of this thesis are the first results of this project. Searches for molecular gas in GPS/CSS have also been carried out \citep[e.g.,][]{O'Dea05}, finding that GPS and large radio sources have similar molecular gas contents and that GPS do not require extremely dense environments. Therefore it is unlikely that the molecular gas will confine the radio source. \\

\item
{\it Interaction with the host}

According to the young source scenario, a powerful radio source will grow from parsec scales to kiloparsec and megaparsec scales. However, to expand to those sizes it must cross the environment of the host galaxy. We therefore expect interaction between the radio lobes and the ISM. The first traces of interaction between GPS/CSS and their hosts were find in the emission line gas. The properties observed in  [\ion{O}{iii}] $\lambda$5007 emission lines \citep{Gelderman94} suggested that, even though the AGN was partly photoionizing the nebula, the radio source was dominating the emission-line kinematics. A few years later, the alignment effect (similar extent for radio and optical wavelengths) seen in large radio sources \citep[e.g.][]{Chambers87, McCarthy87} was found in GPS and CSS \citep{Vries97, Vries99, Axon00}. The alignment between the radio source and emission line gas suggests interaction between them, as the radio source propagates through the host. The emission line gas was found to be brighter in the center of the radio lobes and dimmed with distance. Therefore, the hypothesis adopted was that the gas inside the radio lobe had been shock-ionized while the gas outside was photoionized by the precursor gas in the shock. At the same time, the AGN could contribute to the whole system with photoionization. Cooling time arguments suggested lobe expansion velocities $\gtrsim$1000 \kms \, for most of the sources. In 2002, \cite{O'Dea02} studied the kinematics of the emission line nebulae of CSS sources, finding that the radio source is accelerating the clouds of gas in the host. {\bf Chapter 4} of this thesis continues and wraps up all previous work by studying the ionization mechanisms of three CSS sources. We compare our data with ionization models and confirm that high speed shocks are ionizing the nebula, therefore yielding evidence of jet cloud interaction. {\bf Chapter 5} of this thesis also finds evidence of jet cloud interactions, this time through radio observations of \ion{H}{i} absorption. {\bf Ongoing research} by our group (not included in this thesis) finds a correlation between strength of [\ion{O}{iii}] and radio size, suggesting that the expansion of the radio source is enhancing the [\ion{O}{iii}] emission in the host. Outflows of gas induced by the AGN can also affect the evolution of the host. They can affect even the star formation history of a galaxy \citep[e.g.,][]{Morganti03, Silk98}. Different studies are finding outflows of gas in CSS/GPS sources \citep[e.g.,][and {\bf Chapter 6} of this thesis]{Holt03}. Furthermore, some cases of outflows strongly support the young scenario, as well as interactions between the radio source and the host \cite[e.g.,][]{Tadhunter01, Holt05}. {\bf Chapter 7} of this thesis finds more evidence of radio source -- host interaction, this time through near UV imaging: some CSS sources show jet induced star formation in their hosts.\\

\item
{\it The Starburst--AGN connection and the role of GPS and CSS sources}

It has been suggested that mergers and intergalactic interactions can trigger an AGN event \citep[e.g.,][]{Heckman86, Sanders88, Baum92, Israel98}. Such a strong event can also trigger star formation in a galaxy \citep[e.g.,][]{Ho05}. It is also generally thought that black hole and galactic evolution are related \citep[e.g.,][]{Silk98, Gebhardt00, Begelman05}. However, evidence of a causal connection between starbursts and AGN has yet to be found. Substantial work has been carried out trying to find it in different types of AGN \citep[e.g.,][]{Veilleux01, Gonzalez01, Kauffmann03, Ho05}. Studies of stellar populations in the hosts of GPS, CSS and large (FR I and FR II) sources find traces of episodes of star formation by the time the radio source was formed, or earlier\footnote{This delay is usually consistent with the time needed for gas to be driven into the center \citep[e.g.,][]{Lin88}.} \citep[e.g.,][]{Allen02, O'Dea03, Morganti03b, Raimann05, Johnston05, Tadhunter05, Emonts06}. 

The presumed young ages of GPS/CSS sources makes them perfect tools to study the connection between the starbursts, AGN formation and triggering of the radio source. {\bf Chapter 7} of this thesis presents the first near-UV study of GPS and CSS sources, and looks for evidence for this connection. The results are consistent with a scenario where a strong interaction has triggered a starburst and the formation of the radio source. Studies at different wavelengths have been planned to complement these results.\\

\end{itemize}

\section{Detailed contributions by this thesis}
\label{chapters}

The work carried out in this thesis has contributed to our understanding of GPS and CSS sources and their relation to the host galaxies from different perspectives. In this section I summarize the most important contributions.\\

{\it Improving the samples of known GPS and CSS sources}. Chapters 2 and 3 are mainly devoted to identifying new GPS and CSS sources. Chapter 2 presents Very Large Array (VLA) observations of compact sources in the southern sky. Seven new CSS sources are found and a comparison between the Third Cambridge Catalogue (3C) and Molonglo Southern 4 Jy Sample (equivalent to the 3C in the Southern hemisphere) is carried out. Chapter 3 presents Very Large Telescope (VLT) deep imaging and spectroscopy of GPS candidate sources from the \cite{O'Dea91} master list. Six optical counterparts are identified and four new redshifts are measured. Using data from these chapters and from the literature, five sources are removed from the GPS and CSS lists. Most of them were classified as GPS according to their radio spectra. However, new available data suggest that these sources are variable larger radio sources.\\


{\it Radio morphology of compact sources}. The VLA and the European VLBI Network (EVN) were used to study the morphologies of GPS and CSS sources and to compare them with observations in other wavelengths (Chapters 2, 5 and 6). VLBI radio observations allowed us to localize the atomic hydrogen in these sources and study its relation to the emission line gas. Previous studies suggested that the \ion{H}{i} absorption could be related to the emission line gas and our results seem to confirm it. Using high resolution mapping and previous observations of a long-known GPS source we discovered a previously unseen component, changing our understanding of the source morphology: what it was thought to be a compact double, it may well be a one-sided core-jet source. \\

{\it Interaction with the host}. Chapters 4, 5, 6 and 7 study and search for interaction between the radio source and its host. The main goal is try to understand how the growth of the radio source is affecting the host it lives in, and if the properties of the host affect the evolving radio source. The small size of GPS and CSS sources makes them excellent probes to search for this interaction. Their radio lobes are still inside the host so the interaction between them is expected to be higher than in larger radio sources. We find evidence of this interaction. Our Hubble Space Telescope (HST) Space Telescope Imaging Spectrograph (STIS) spectroscopic data (Chapter 4) reveals that the shocks produced by the expanding radio source are ionizing (in some sources almost completely) the emission line gas of the host. Our EVN radio observations find \ion{H}{i} associated with the emission line gas and outflows, supporting the interaction scenario. Using the HST Advanced Camera for Surveys we carried out the first near-UV study of GPS/CSS sources. The high spatial resolution of our images unveils complex kpc scale morphologies in the hosts of these sources. Color information is still needed. However, our first results suggest recent star formation episodes in the hosts on time scales consistent with the triggering of the radio source. Some sources show radio and UV structures that suggest star formation triggered by the expansion of the radio lobes through the host.\\


{\it Star formation and nuclear activity}. Using our optical and ultraviolet images (Chapters 3 and 7), we studied the stellar population properties of the hosts of GPS and CSS sources. The lack of colors make our results preliminary. However, these results are found to be consistent with previous studies of the stellar populations in GPS and CSS sources. Comparison with synthetic stellar population models suggest that GPS and CSS sources occur in galaxies with recent star formation and  suggest a connection between nuclear activity and starburst. Furthermore, the UV images seem to imply that the expansion of the radio source can affect the star formation history of the host.\\


\section{The bottom line: Strong interaction between radio source and host}
\label{bottomline}

The main goal of this thesis is to study the interrelation of powerful radio sources with their hosts. The objects of study are GPS and CSS sources. Due to their small size, GPS/CSS sources are excellent probes of this relation. Furthermore, their young age allows us to compare them to the larger, old radio sources and establish a time-line evolution of this relation. \\

This thesis combines imaging and spectroscopy of GPS/CSS sources at different wavelengths, and all our studies lead to the same conclusion: the presence and expansion of powerful radio sources clearly affect the properties and evolution of their hosts. All chapters of the thesis (excluding the sample studies) find evidence of strong interaction between the host and the radio source. Furthermore, the radio source and host can significantly affect each others  evolution. However, this influence takes place in different ways. The influence that the host has on the radio source is somehow indirect. However it can completely change its destiny: depending of the contents, distribution and density of the gas, the radio source will die early, expand and grow into the large FR sources, or remain confined inside its host. In contrast, the influence of the radio source on its host seems to be more direct and takes place during its expansion through the host: the radio source will affect the kinematics and ionization of the emission line gas, and may change the star formation history of the host.\\

Briefly, the main results that lead to this conclusion are:
\begin{enumerate}

\item
Presence of shock ionized gas in 3C~67, 3C~277.1 and 3C303.1 (Chapter 4).

\item
\ion{H}{i} gas associated to the emission line gas in 3C~49 and 3C 268.3 and the presence of outflows in B2050+364 (Chapters 5 and 6).

\item
Possible events of star formation related to the triggering of the radio source and findings of jet induced star formation in 1814--637 and 3C 303.1, which is also the source showing the strongest contribution from shocks to the ionization of the emission line gas (Chapter 7).
\end{enumerate}



\section{Ongoing and future work}
\label{future}

Much progress has been made in understanding the nature of powerful radio galaxies and GPS/CSS sources, as well as their relation to their hosts. However, there are still some puzzling questions to be answered. A few of these are clear follow-ups of the issues listed above (Section \ref{developments}). Others may affect more general subjects such as galaxy formation and evolution, i.e.  why do some galaxies harbor an AGN? Is the AGN phenomenon a phase all or just certain galaxies go through? What are the conditions needed in a galaxy to host an AGN? \\

The younger the radio source, the closer we are to the formation of the AGN. Therefore, GPS and CSS sources may play an important role in answering some of these questions. However, GPS and CSS sources have unanswered questions of their own. In this section I will describe some of the ongoing work we are carrying out and suggest possible lines of continuation of the research carried out in this thesis.\\ 


Large, homogeneous samples are needed to test models, solve the GPS quasar problem, test scenarios, etc. Considerable amounts of work are being carried out aiming to re-address GPS/CSS samples and obtain new ones. However, there is still much to be done, especially for the smallest and unresolved sources. High resolution VLBI imaging is needed, as well as variability studies. Searches for extended emission are also useful as they can give an estimate of how many GPS/CSS sources are recurrent or have been interrupted. The classification of the GPS candidates in the  \cite{O'Dea91} master list is almost complete. VLT searches for optical counterparts and redshifts have proven useful. VLT images and spectroscopy will be carried out for the remaining objects still lacking identification. Measuring radio fluxes in some of these sources will help finding variable sources and sort out incorrect classifications.\\

Studying the gas contents and distribution will continue to constrain issues such as young versus old scenario or the free-free versus synchrotron self-absorption problems. \cite{Vermeulen03} find \ion{H}{i} absorption in one third of their sources. Chapters 5 and 6 yielded interesting results in both the distribution of the \ion{H}{i} gas and the morphology of three of these GPS/CSS sources. Therefore, we expect that EVN observations of the remaining sources will also yield interesting results. In the same line of research, observations searching for molecular gas are not very frequent in GPS and CSS sources and will be planned. Techniques such as measuring the polarization properties and Faraday rotation are also useful in the study of the gas polarization of the hosts of GPS and CSS sources. X-ray observations are also yielding interesting results on gas contents, especially close to the nucleus, as well as on the strength of the AGN.\\ 



GPS and CSS are the progenitors of large radio sources.  However, it is not clear how their expansion takes place. Theoretical models keep being developed. However, they are still too simple and cannot accurately reproduce the observations. Considerable amounts of work are still needed both observationally and theoretically to be able to reproduce in detail the expansion of the radio lobes, how they interact with the host, etc. We are carrying out a comparison of these models, as well as jet expansion models with a large collection of GPS/CSS and FR sources aiming to constrain the time evolution of the radio source and check the predictions from the different models.\\

Our ACS images are consistent with a connection between a recent starburst and the formation of the radio source. However, we lack color information to accurately establish how much of the light is from young stars. The new and soon to be launched IR space telescopes (Spitzer, ASTRO-F, HERSCHEL) provide the perfect tools for obtaining high resolution images and spectral information on the stellar composition (as well as dust and gas) of the hosts of GPS and CSS sources.\\

In a longer time span, the improvements in our radio observing capabilities that the Square Kilometer Array will provide will allow us to study much weaker AGN, as well as obtain information on contents and distribution of gas in the most inner regions on the AGN. In the infrared, the high resolution and extremely high sensitivity of the James Webb Space Telescope, combined with the Integral Field Spectroscopy unit of the Mid Infrared Instrument will allow us to get extremely detailed 2D maps of the properties and distribution of the emission line gas in hosts of AGN. Furthermore, it will permit such studies, not only in the nearby universe but also at high redshifts. We may be able of tracing the cosmological evolution of the interaction of powerful radio sources and their host, out to the time the first galaxies were forming.\\

 \markboth{}{}	

\bibliographystyle{apj}						
\bibliography{ALORefs}
\addcontentsline{toc}{chapter}{Bibliography}
\markboth{Bibliography}{Bibliography}


\chapter*{List of Publications}
\addcontentsline{toc}{chapter}{List of Publications}
\markboth{List of Publications}{List of Publications}

\sloppy
\thispagestyle{empty}

\normalsize

\noindent
{\large{\bf Publications in refereed journals}}~: \\

\noindent
{\it \ion{H}{i} absorption in 3C49 and 3C268.3. Probing the environment of Compact Steep Spectrum and 
      GHz Peaked Spectrum sources}\\
 Labiano, A., Vermeulen, R., Barhel, P.D., O'Dea, C.P.,  Gallimore, J F.,  Baum, S.A., de Vries, W. H.\\ A\&A. In press. Astro-ph/0510563\\

\noindent
{\it Atomic hydrogen in the one-sided 'compact double' radio galaxy 2050+364}\\
 Vermeulen, R., Labiano, A., Barthel, P.D., Baum,  S.A., de Vries, W. H.,  O'Dea, C.P. \\ A\&A. In press. Astro-ph/0510440\\

\noindent
{\it HST/STIS low dispersion spectroscopy of three Compact Steep Spectrum sources. Evidence for 
     jet-cloud interaction}\\
 Labiano, A., O'Dea, C. P., Gelderman, R., de Vries, W. H., Axon, D. J., Barthel, P. D., Baum, S. A., 
     Capetti, A., Fanti, R., Koekemoer, A. M., Morganti, R., Tadhunter, C. N., 2005 A\&A, 436, 439.\\

\noindent
{\it HST/STIS spectroscopy of CSS sources. Kinematics and ionization of the aligned nebulae}\\
 Labiano, A., O'Dea, C. P., Gelderman, R., de Vries, W. H., Axon, D. J., Barthel, P. D., Baum, S. A., 
     Capetti, A., Fanti, R., Koekemoer, A. M., Morganti, R., Tadhunter, C. N.\\
 Publications of the Astronomical Society of Australia 2003, 20, 28.\\

\noindent
{\large{\bf Other publications}}~: \\

\noindent
{\it Star formation in hosts of young radio galaxies}\\
Labiano, A., O'Dea, C.P., Barthel, P.D., de Vries, W.H., Baum, S.A. \\
New Astronomy Reviews. 2005. In press. Astro-ph/0512057\\

\noindent
{\it Spectroscopy of CSS sources}\\
Labiano, A., O'Dea, C. P., Gelderman, R., de Vries, W. H., Axon, D. J., Barthel, P. D., Baum, S. A., 
    Capetti, A., Fanti, R., Koekemoer, A. M., Morganti, R., Tadhunter, C. N., \\
 Highlights of Spanish Astrophysics, 2002, Volumen 3, 219.\\

\noindent
{\it Radio Astronomy at the Robledo Deep Space Stantion}\\
G\'omez, J. F., Garc'a-Mir\'o, C., Labiano, A., Alberdi, A.\\
   Highlights of Spanish Asttrophysics, 2001, Volumen 2, 377. \\

\noindent
{\it A stellar library of H and He line absorption profiles at high resolution}\\
D\'iaz, A. I., \'Alvarez, M., Moll\'a, M., Labiano, A., Gonz\'alez Delgado, R. M., P\'erez, E., V\'ilchez, J.M\\
  Highlights of Spanish Asttrophysics, 2001, Volumen 2, 105. \\


\chapter*{Acknowledgements}
\addcontentsline{toc}{chapter}{Acknowledgements}
\markboth{Acknowledgements}{Acknowledgements}

\vspace{-0.5cm}

Along my graduate life I have had the chance of meeting an incredible number of people who have supported me in one way or another. \\

The list is long so I will start from the beginning. Don Pepe y todos esos profesores que han aguantado a este empoll\'on con sue\~{n}os de astr\'onomo desde la infancia. The astronomy department in the Universidad Auton\'oma de Madrid, who hosted me since '96! Gracias por confiar en mi, \'Angeles. El Seminario de Ciencias Planetarias de la Complutense. Gracias por acoger f\'isicos, Paco! The priceless LAEFF family! Rosa, Enrique y Luis, que de una u otra forma me hab\'eis animado desde el principio, y segu\'is haci\'endolo! The 2001 and 2005 generations of {\it Kapteyners} who welcomed me and will see me off. I am sorry we have not had the chance to meet better. And, of course, the Space Telescope Science Institute, a dream came true on January 8, 2002.\\


I am particularly grateful to my supervisors Chris O'Dea and Peter Barthel. Thank you for believing in me from the beginning, your help, and especially your support and patience trying to get a scientist out of this student. Thanks also to your families for their hospitality on countless occasions! \\

And last but by no means least, much love to my Family and Friends. You have always been there for me, even through the worst times. I will not let you down.\\

\begin{center}
My most sincere thanks to all of you.\\
\end{center}

\begin{flushright}
{\it \'Alvaro Labiano Ortega \\ Groningen, December 31, 2005}
\end{flushright}

 \markboth{}{}		



\end{document}